\documentclass[pra,aps,showpacs,twocolumn,letterpaper]{revtex4}
\usepackage{epsf}
\usepackage[dvips]{color}
\newcommand {\be}{\begin{equation}}
\newcommand {\ee}{\end{equation}}
\newcommand {\bea}{\begin{eqnarray}}
\newcommand {\eea}{\end{eqnarray}}

\begin{document}

\title{A pulsed atomic soliton laser}
\author{L.~D. Carr}
\affiliation{JILA, National Institute for Standards and Technology and University of Colorado, Boulder, CO 80309-0440}
\author{J. Brand}
\affiliation{Max Planck Institute for the Physics of Complex
Systems, N\"othnitzer Stra{\ss}e 38, 01187 Dresden, Germany\\}

\date{\today}

\begin{abstract}
It is shown that simultaneously changing the scattering length of
an elongated, harmonically trapped Bose-Einstein condensate from
positive to negative and inverting the axial portion of the trap,
so that it becomes expulsive, results in a train of self-coherent
solitonic pulses.  Each pulse is itself a non-dispersive
attractive Bose-Einstein condensate that rapidly self-cools. The
axial trap functions as a waveguide.  The solitons can be made
robustly stable with the right choice of trap geometry, number of
atoms, and interaction strength.  Theoretical and numerical
evidence suggests that such a pulsed atomic soliton laser can be
made in present experiments.
\end{abstract}


\maketitle

\section{Introduction}

Solitons have applications in a wide variety of physical contexts,
ranging from water waves to photonic
crystals~\cite{hasegawa1,drazin1,agrawal1}. For example, they have
been used in transatlantic communications systems, where the need
for expensive amplifiers mid-line in fiber optic cables that run
over long distances is reduced or eliminated.  By virtue of their
many uses, as well as their mathematical beauty, solitons are a
continuing subject of vigorous research (see, for
example,~\cite{kivshar3,sulem1}).  In particular, they have proven
highly useful in laser applications~\cite{agrawal1}. Nonlinear
materials are used to cause high-intensity coherent light waves
emitted by lasers to self-focus into stable non-dispersive pulses.

Bose-Einstein condensates (BEC's) are coherent matter waves in
analogy to coherent light waves. BEC's are usually generated as a
standing wave in a trap which functions as a cavity. When
outcoupled from the trap, a BEC can provide a highly brilliant
source of coherent matter-wave radiation, and as such is commonly
called an \emph{atom laser}.
The challenge in
making a BEC into a useful atom laser is in the outcoupling~\cite{mewes1,hagley1}.
To this end, many experimental
methods have been developed.  Anderson and Kasevich~\cite{anderson1998} tilted a
BEC trapped in a periodic potential created by a standing light wave.  The gravitational field
induced by the tilt caused the condensate to tunnel through the wells and interfere coherently, thereby
creating a pulsed atom laser.  Bloch et al.~\cite{bloch2}
used an external laser to change the spin state of atoms in two locations in a harmonic magnetic trap.
The condensed atoms then spilled out, again due to gravitational effects;
the two separate outcouplings allowed them to make the
first clear demonstration of coherence along the whole length of the beam.  Many experiments
and proposals have followed.  All of these atom lasers were studied in the context of
repulsive BEC's.  Repulsive BEC's naturally disperse in all directions; even axial confinement
in a waveguide cannot prevent spreading in the direction of propagation.

The unique contribution of the present study is to show how an
\emph{attractive} BEC can be made into an atom laser.  A repulsive
BEC fractures near conducting surfaces~\cite{leanhardt2002},
spreads out, and in general is easily excited. In contrast, an
attractive BEC, so long as it is axially confined and the
experimental parameters are chosen properly, can be made into a
pulsed atomic soliton laser which is robustly stable against all
of these effects. Moreover, attractive BEC's in this form may be
superior to repulsive BEC's in applications to atom
chips~\cite{ott2001,hansel2001} and
interferometry~\cite{kasevich2002}. For instance, in the
``nevatron'', a BEC storage ring, the wave packet of repulsive
Bose-condensed atoms circulates a few times before spreading out
and decohering~\cite{sauer2001}. This effect is accentuated by
superconducting wires which lie transverse to the direction of
propagation. A bright soliton would not only be non-dispersive,
but, even if excited by the passage over the jump in potential
created by the wire, would quickly self-cool by emitting a small
fraction of its atoms, typically less than a fraction of a
percent~\cite{satsuma1,carr30}. It could therefore circulate
indefinitely, subject to three-body recombination rates and other
effects beyond those of the mean field~\cite{dalfovo1}.

{ In the following, we explain how to create such a pulsed atomic
soliton laser.  Bright matter-wave solitons have been created,
both singly~\cite{carr29} and in trains~\cite{strecker1}. It is
shown that a combination of the experimental techniques of
Refs.~\cite{carr29} and~\cite{strecker1}, together with the right
choice of parameters, suffices to create a pulsed atomic soliton
laser from an attractive BEC. After presenting the basic method in
Sec.~\ref{sec:basic}, we illustrate its viability via
three-dimensional simulations in Sec.~\ref{sec:sims}. Then, in
Secs.~\ref{sec:criteria} and~\ref{sec:feature}, the stability
criteria and important dynamical features of the pulsed atomic
soliton laser are explained in detail. In Sec.~\ref{sec:params},
the simulations are discussed in light of the results of
Secs.~\ref{sec:criteria} and~\ref{sec:feature}.  Finally, in
Sec.~\ref{sec:conclude} we conclude.}

\section{Basic Method}
\label{sec:basic}

The 3D Nonlinear Schr\"odinger equation (NLS) or Gross-Pitaevskii
equation which describes the mean field of the BEC is written
as~\cite{dalfovo1}
\begin{equation}
\label{eqn:gpe3d} \left[ -\frac{\hbar^2}{2m} \nabla^2
+g\,N\,|\Psi|^2 +V(\vec{r}) \right]\Psi =i\hbar\partial_t\Psi \, ,
\end{equation}
where \be V(\vec{r})\equiv \frac{1}{2}m(\omega_{\rho}^2\rho^2
+\omega_z^2 z^2) \, ,\ee $g\equiv 4\pi\hbar^2 a/m$, $a$ is the
$s$-wave scattering length,
$m$ is the atomic mass, $N$ is the number of condensed atoms, the
condensate order parameter $\Psi=\Psi(\vec{r},t)$ has been
normalized to one, and axisymmetric harmonic confinement has been
assumed. Note that for negative scattering length, or attractive
nonlinearity, solutions are liable to collapse in certain
parameter regimes~\cite{ruprecht1,sulem1}, as shall be discussed
below. With the exception of Sec.~\ref{ssec:decohere}, where the
decoherence time between pulses is estimated, it is assumed that
the BEC is described by Eq.~(\ref{eqn:gpe3d}).

The basic method for creating the pulsed atomic soliton laser is
as follows.

\begin{enumerate}
    \item A repulsive BEC is created in an elongated harmonic trap such
that $\omega_z\ll\omega_{\rho}$.
    \item The scattering length is made small and negative via the now
well-established experimental technique of using a magnetically
induced Feshbach
resonance~\cite{vogels1,inouye1,carr29,strecker1}.
    \item Simultaneously, the axial potential is changed from small and
attractive ($\omega_z$ real) to small and expulsive ($\omega_z$
imaginary)~\cite{carr29,carr30,abdullaev2003}.
    \item The condensate becomes modulationally unstable to
spatial pulse formation. This instability is non-dissipative.
The pulses are seeded by self-interference of the order parameter,
as we have elsewhere described~\cite{carr34}. The initial growth rate of
the modulational instability can be calculated via linear
perturbation theory~\cite{hasegawa2}.
    \item The ensuing solitonic pulses are subject to {\it primary collapse}
in two or three dimensions, as well as {\it secondary collapse}
due to soliton--soliton interactions~\cite{carr34}. Furthermore,
if they are too long in the $z$-direction they can be torn apart
by the force of the expulsive harmonic potential~\cite{carr30}.
With the right choice of parameters, these effects can be avoided
and the solitonic pulses made robustly stable, as shall be
described in Sec.~\ref{sec:criteria}.
    \item The solitons continue to accelerate.  Their relative spacing
increases as $\Delta z \propto \exp(|\omega_z|\, t)$.
    \item The
expulsive harmonic potential can eventually be coupled to a linear
or even a flat potential for applications. We do not discuss the
various possibilities here.
\end{enumerate}

\begin{figure}[t]
\epsfxsize=8cm \leavevmode  \epsfbox{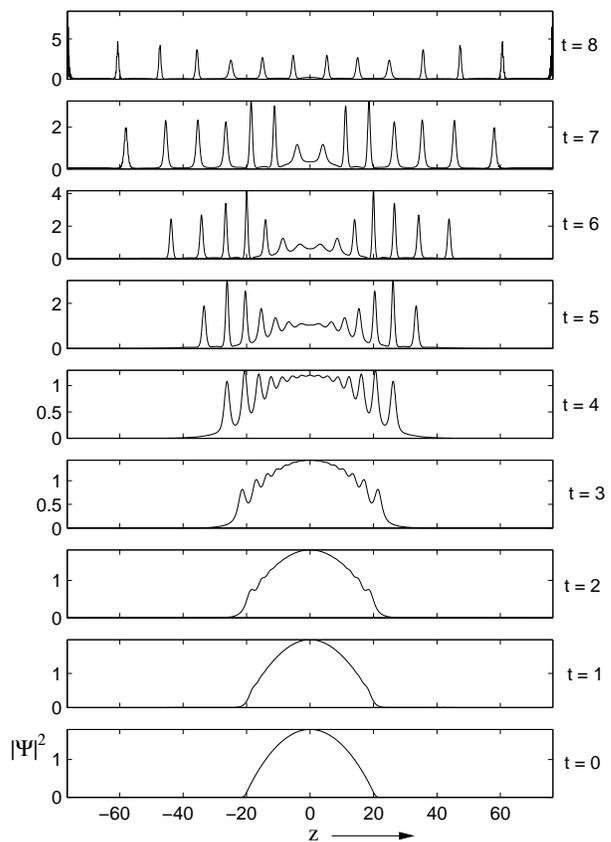}
\caption{\label{fig:1}  Shown is the evolution of an attractive
Bose-Einstein condensate into a pulsed atomic soliton laser. Time
slices of the line density in $z$ are shown for $x,y=0$.
Modulational instability of the initial density profile is seeded
by self-interference of the order parameter, so that solitons form
first at the cloud edges and later towards the center.  The
latest, top panel, shows that a well-separated set of stable
solitonic pulses are produced. Note that, for $N=10^4$ atoms,
$a=-3 a_0$, and a trap geometry of $\omega_{\rho}=2\pi\times 2.44$
kHz, { $\omega_{z}=2\pi i\times 2.26$ Hz, the time units are
scaled to 22 ms and the spatial units to 10 $\mu$m.}}
\end{figure}

\begin{figure*}[t]
\epsfxsize=17cm \leavevmode  \epsfbox{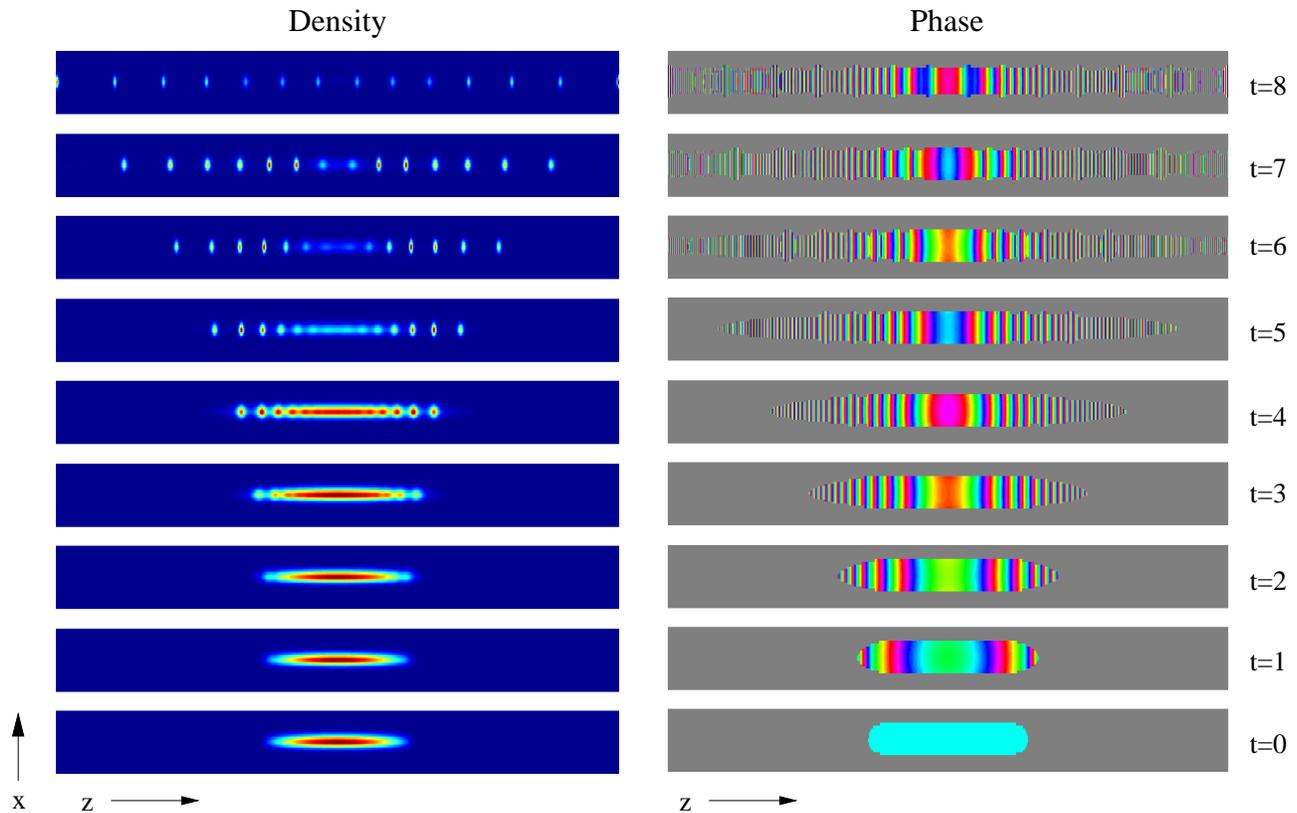}
\caption{\label{fig:2} Shown are the evolution of the density and
phase along a two dimensional cut at $y=0$ for the simulation of
Fig.~\ref{fig:1}.  A set of well-defined solitonic pulses is
evident in the latest (top) panel.  The strong variations in the
phase at late times is due to the high momentum of the solitons
caused by the expulsive harmonic potential. Note that the phase is
shown on the color circle, {\it i.e.}, modulo $2\pi$, while the
density is in arbitrary relative units rescaled for each plot. The
aspect ratio of the plots showing a region of 0.822 by 153 length
units was changed for visualization; length and time units are the
same as in Fig.~\ref{fig:1}.}
\end{figure*}

An important point is that even if the coupling is not smooth,
each soliton responds to perturbation by a shift in its phase and
by emitting a small number of atoms, typically a fraction of a
percent of the total number in the soliton. Insofar as an excited
BEC described by the NLS models a condensate plus thermal
field~\cite{sinatra2001,davis2002a}, where the ``condensate'' is a
stationary solution, one may term this process {\it self-cooling}.
As each soliton is itself a BEC, this model can be applied to each
pulse separately.  Self-cooling to $T=0$ in an expulsive harmonic
potential occurs exponentially, with the density fluctuations at
the center of each solitonic pulse falling off as
$\exp(-|\omega_z| \,t)$~\cite{carr30}.

One may ask what this schemata has in common with the operation of
a normal, light-wave laser. Insofar as the initial condition is an
excited mode of the harmonic trap plus mean field potential, and
the ensuing pulse train is a much lower energy mode, this
situation has a certain analogy with population inversion.   As
was mentioned in the introduction, the initial trapping potential
may be considered as a cavity, with the outcoupling provided by
the sudden change to an expulsive trapping potential in the axial
direction.  However, the emission of solitonic pulses is not
stimulated, as strictly required for the use of the acronym LASER
(light amplification by stimulated emission of radiation), but
rather spontaneous. The analogy of the proposed scheme to the
operation of a laser is therefore rather in the output than in the
detailed mechanism of its operation: one produces a train of
non-phase-locked self-coherent pulses.

\section{Pulsed atomic soliton laser dynamics: Proof of principle simulations}
\label{sec:sims}

Three dimensional simulations of Eq.~(\ref{eqn:gpe3d}) were
performed, with parameters which satisfied the criteria given in
Sec.~\ref{sec:criteria}. Cylindrical symmetry was assumed in order
to make computations with a large grid size possible ($2048\times
16$) \cite{ref:laguDVR}. The initial profile was obtained by
imaginary time relaxation. This resulted in a Thomas-Fermi-like
profile in the $z$ direction (see Eq.~(\ref{eqn:tfprofile}) below
and~\cite{dalfovo1}), and a nearly Gaussian one in $x$ and $y$. A
trap of aspect ratio $\omega_\rho / \omega_{z0} =
(\ell_{z0}/\ell_\rho)^2 = 538$ and nonlinearity parameter $a_{s0}
N/\ell_{z0} = 1.02$ was used to produce the initial state, where
$\omega_{z0}$, $\ell_{z0}$ and $a_{s0}$ all refer to these initial
conditions, and $\ell_i\equiv \sqrt{\hbar/(m\omega_i)}$.

At $t=0$, the longitudinal trapping frequency $\omega_{z0}$ was
changed to a weak expulsive harmonic potential with $\omega_z =
0.5 i \,\omega_{z0}$, and the nonlinearity was switched from
repulsive to attractive, with $-a N/\ell_{z0} = 0.0854 = 0.00368\,
\ell_{z0}/\ell_\rho$. As attractive BEC's can collapse in three
dimensions, this is an important point in the choice of
parameters.  The length unit in the simulations is $u = 0.56
\ell_{z0}$ and the time unit $\tau = 2 m u^2 /\hbar$. To compare
with experimental parameters, one must choose the number of
particles and a scaling factor, e.~g.~$u = 10 \mu m$ and $N=
10^4$, which corresponds to
 $\tau = 22ms$, $\omega_{z0} = 2\pi \times 4.53$Hz, $\omega_z =
2\pi\,i \times 2.26$Hz, and $\omega_\rho = 2\pi \times 2.44$kHz
with a scattering length of $a = -3 a_0$.

Figures~\ref{fig:1} and~\ref{fig:2} illustrate the evolution of
the density and phase of the condensate in time slices through the
$x$-$z$ plane for $y=0$ with the above described initial state and
parameters.  Several observations may be made based on the
figures.  { Firstly, the final number of solitons is
14.  Secondly, they are stable against collapse and, once formed,
do not subsequently interact over the lifetime of the simulation.
Thirdly, solitons form first at the edges of the cloud, then later
towards the center, as was also observed in a simplified model in
our previous work~\cite{carr34}. We note that no white or colored
noise was added to this simulation.  The reasoning behind our
choice of parameters will be discussed in Sec.~\ref{sec:criteria},
while the details of the simulation itself will be interpreted in
Secs.~\ref{sec:feature} and~\ref{sec:params}.}

\section{Stability criteria}
\label{sec:criteria}

 In order that the pulsed atomic soliton laser
be robustly stable over the lifetime of an experiment, a set of
criteria must be satisfied.  These criteria are detailed in the
following subsections.  Note that in the below considerations we
are interested in stability for experimental purposes, not
mathematical stability to infinite time.

\subsection{Two-dimensional primary collapse}
\label{ssec:2dcollapse}

Two-dimensional transverse primary collapse must be avoided. In
the case of strongly anisotropic axisymmetric confinement, one may
adiabatically separate the slow longitudinal from the fast
transverse degrees of freedom. The adiabatically varying
transverse state obeys a 2D NLS which shows an instability towards
collapse. The criterion for stability found by numerical
integration of the 2D NLS is \be\label{eqn:transv} -8 \pi a
n_{{\rm 1D}}(z,t) < \eta_c^{{\rm 2D}}=11.7\ldots \, , \ee where
$n_{{\rm 1D}}(z,t)$ is the local axial line density of the
condensate~\cite{weinstein1983}. If adiabaticity is violated,
collapse can also happen at weaker nonlinearity due to transverse
oscillations on a time scale
$\pi/\omega_{\rho}$~\cite{pitaevskii1996}. When the longitudinal
dynamics is significantly slower than this time scale, the
adiabatic separation of scales is valid. If, additionally, the
transverse nonlinearity is weak, {\it i.e.}, $|8 \pi a n_{{\rm
1D}}(z,t)| \ll 1$, the longitudinal equation reduces to the
quasi-1D NLS
\be \left[-\frac{\hbar^2}{2m}\partial_z^2
+g_{{\rm 1D}} N \left|\psi\right|^2 + \frac{1}{2}m\omega_z^2 z^2
\right]\psi=i\hbar\,\partial_t\,\psi \,
,\label{eqn:gpe1d}\ee
where $g_{{\rm 1D}}\equiv 2\,a\,\omega_{\rho}\hbar$ is the
renormalized quasi-1D coupling constant~\cite{carr30}, provided
$\ell_{\rho}\gg |a|$~\cite{olshanii1}, with
$\ell_{\rho}\equiv\sqrt{\hbar/(m\omega_{\rho})}$.

\subsection{Three-dimensional primary collapse}
\label{ssec:collapse3d}

Three-dimensional primary collapse of the individual solitonic
pulses in the atom laser must be avoided.  The static condition
based on imaginary time relaxation of Eq.~(\ref{eqn:gpe3d})
is~\cite{carr30} \be -\frac{N_{\mathrm{ais}} a} {\ell_{\rho}} <
\eta_c^{{\rm 3 D}} = 0.627\ldots\, , \label{eqn:collapse3d} \ee
where $N_{\mathrm{ais}}\sim N/N_s$ is the number of atoms in a
soliton, with $N_s$ the number of solitons (see
Sec.~\ref{sec:params} below). Replacing the $<$ sign with a $\ll$
sign ensures stability for an excited soliton.  Note that the
value of $\eta_c^{{\rm 3 D}}$ can vary slightly as the anisotropy
of the trap changes~\cite{carr30,gammal1}. However, 3D collapse is
essentially an isotropic phenomenon, with the trap simply setting
the initial conditions~\cite{ng2004}.

\subsection{Explosion of individual solitonic pulses}
\label{ssec:explode}

The soliton can become unstable when the expulsive potential
overcomes the balance between the mean field energy and kinetic
energy necessary for the soliton's existence and destroys it by
tearing it in two~\cite{carr30}. We term this kind of possible
instability \emph{explosion}. In order to avoid explosion, the
geometry must be chosen so that $\ell_{z} \gg
\ell_{\mathrm{sol}}$, where $\ell_z\equiv
\sqrt{\hbar/m|\omega_z|}$. The soliton length
$\ell_{\mathrm{sol}}\sim \xi$, where $\xi$ is the healing length,
can be determined as follows. Taking the form of the soliton as
the well-known solution in one dimension for a constant potential
(see~\cite{carr30} and references therein) \be
\psi(z,t)=\frac{1}{\sqrt{2\ell_z}}\,\mathrm{sech}
\left(\frac{z}{\ell_z}\right)e^{-i\mu t}\, ,\ee and substituting
into Eq.~(\ref{eqn:gpe1d}) while temporarily neglecting the
trapping potential, one obtains
$\ell_z=2\hbar^2/(m|g_{\mathrm{1D}}|N_{\mathrm{ais}})$. Then \be
\ell_{\mathrm{sol}} \simeq
\frac{\ell_{\rho}}{|a|{N_{\mathrm{ais}}}}\ell_{\rho} \, ,\ee where
the wavefunction has been renormalized to account for the division
of $N$ total atoms into $N_{\mathrm{ais}}$ atoms in any given
soliton.  Note that the prefactor is the inverse of the ratio
which must be small to avoid 3D collapse, as given by
Eq.~(\ref{eqn:collapse3d}) in the preceeding subsection.

A variational analysis based on a hyperbolic secant {\it ansatz}
and Eq.~(\ref{eqn:gpe1d}) with the trapping potential gives a more
precise condition to avoid explosion as~\cite{carr30} \be
\frac{\ell_z}{\ell_{\mathrm{sol}}}
> \left(\frac{2^6\pi^4}{3^3}\right)^{1/4} =
2.20\ldots\, .\label{eqn:noexplode}\ee  In the quasi-1D regime far
from collapse, such a variational analysis typically gives
estimates to better than 1\%.  As in Eq.~(\ref{eqn:collapse3d}),
the $>$ sign can be replaced with a $\gg$ sign to ensure stability
for an excited soliton.

\subsection{Soliton--soliton interaction and secondary collapse}
\label{ssec:interaction}

The harmonic potential must be sufficiently strong so as to
prevent secondary collapse caused by soliton--soliton interaction.
If two solitons overlap coherently they can violate the stability
criterion of Eq.~(\ref{eqn:collapse3d}), due to the doubling of
the number of atoms.  Even if, due to decoherence during soliton
propagation (see Sec.~\ref{ssec:decohere} below), their relative
phase is not defined prior to interaction, upon interacting they
develop a well-defined relative phase~\cite{cct1999}.

In order to treat soliton motion in a slowly varying potential, as
defined explicitly by the condition of Eq.~(\ref{eqn:noexplode}),
one may suppose a separation of scales, as may be formally defined
by a multiscale analysis (see, for example,
Ref.~\cite{michinel1}). In this case an approximate equation of
motion for the relative soliton parameters is given
by~\cite{gordon2,elyutin1} \bea
\ddot{\phi} &=& \frac{8\hbar^2 \kappa^4}{m^2}\sin(\phi)\exp(-\kappa r)\nonumber\\
\ddot{r}&=& - \frac{8\hbar^2 \kappa^3}{m^2}\cos(\phi) \exp(-\kappa
r) - \omega_z^2 r\, , ~\label{eqn:2soliton} \eea where \be
\kappa\equiv -N g_{\mathrm{1D}}m/\hbar^2\, , \ee $\phi\equiv
\phi_1 - \phi_2$ is the relative phase, $r\equiv |z_1 - z_2|$ is
the relative position, with the two solitons indicated by indices
$1,2$, and motion according to Eq.~(\ref{eqn:gpe1d}) has been
assumed. Equation~(\ref{eqn:2soliton}) describes a separated
soliton pair, {\it i.e.}, the motion outside the interaction
region: it breaks down as they overlap fully.  With respect to
Refs.~\cite{gordon2,elyutin1}, we have here added the physical
units relevant for the BEC and the expulsive harmonic potential.

To prevent soliton--soliton interaction it is necessary that the
potential due to the expulsive harmonic potential be much stronger
than that due to the attraction between solitons.  Taking $\phi(t)
\equiv \dot{\phi}(t)\equiv 0$, which assumes that the solitons are
initially in-phase and have the same amplitude, the two potentials
are given by
\bea V_{\mathrm{ho}}=\frac{1}{2}m\omega_z^2 r^2 \, ,\\
V_{\mathrm{sol}}=8\frac{\hbar^2\kappa^2}{m^2} \exp(-\kappa r) \, .
\label{eqn:forces}\eea In case the solitons are not initially in
phase or do not have the same amplitude, the criterion will only
be less stringent. It is therefore sufficient that {
\be \left(\frac{\ell_{\rho}}{\ell_z}\right)^4
\left(\frac{\ell_{\rho}}{|a|N_{\mathrm{ais}}}\right)^4
\frac{N_{\mathrm{ais}}^2}{N_s^2} \gg \frac{2^6}{\pi^2}\exp(-8\pi
N_s/N_{\mathrm{ais}}) \, .\label{eqn:nointeract}\ee }Here the
first factor in parentheses is the trap aspect ratio while the
second factor is again the inverse of the 3D collapse criterion of
Eq.~(\ref{eqn:collapse3d}).

The question then arises as to whether or not the trap can be made
sufficiently strong so as to prevent soliton interactions, as
required by Eq.~(\ref{eqn:nointeract}), and at the same time
sufficiently weak so as not to cause the individual solitons to
explode, as required by Eq.~(\ref{eqn:noexplode}).  Putting these
two criteria together, one finds \be \chi^{-2} \gg
\frac{3^3}{\pi^6}\exp(-8\pi\chi)\, ,\ee where { $\chi
\equiv N_s/N_{\mathrm{ais}}$}. This relation is always fulfilled,
showing that the two criteria are compatible.

\section{Dynamical features}
\label{sec:feature}

There are two aspects of the dynamics which are necessary to
discuss in detail.  Firstly, there has been some debate as to the
mechanism of soliton formation.  In Sec.~\ref{ssec:interfere}, it
is argued that both dynamically generated fluctuations from
self-interference of the order parameter~\cite{carr34} and noise
due to thermal fluctuations~\cite{khawaja2002} or fluctuations in
the trapping potential~\cite{carr22}, as seen experimentally close
to surfaces~\cite{fortagh2002}, cause the BEC to become
modulationally unstable on approximately the same time scale.
Secondly, outside of the mean field model encapsulated in the NLS,
one may ask how long it takes for the relative phase of solitonic
pulses in the atom laser to randomize, or decohere. In
Sec.~\ref{ssec:decohere}, an estimate of this time scale is made.

Two issues which we do not discuss in any detail are quantum
evaporation and center of mass motion.  The former is studied in
detail in Ref.~\cite{carr30}, where it is shown that matter-wave
bright solitons in an expulsive potential evaporate and eventually
explode.  However, the tunneling rates are so small in the
parameter regimes of interest to the present work so as to be
unimportant. With regards to the latter, in a harmonic potential
the center of mass and relative degrees of freedom are entirely
decoupled, so that we need only consider the relative soliton
motion~\cite{garciaripoll2}.  The center of mass motion is, in any
case, trivial:
$z_{\mathrm{com}}(t)=z_{\mathrm{com}}(0)\exp(|\omega_z| t)$ in the
quasi-1D regime.

\subsection{Seeding of modulational instability: self-interference vs. noise}
\label{ssec:interfere}

In order to understand the mechanism of modulational instability
for a non-uniform initial density profile and in the presence of a
non-constant potential, it is necessary to briefly review
modulational instability in the uniform case, which is well known
from fiber optics~\cite{hasegawa2}. A linear response analysis
reveals that, for attractive nonlinearity, a small sinusoidal
modulation of a uniform state $\psi_0$ with wavenumber $k$ grows
{with time} at a rate $\gamma$ given by \be
\gamma^2=-\frac{\hbar^2}{4m^2} \left[k^2-\frac{2 m
|g_{\mathrm{1D}}| \, n_{{\rm 1D}}}{\hbar^2}\right]^2
+\frac{n_{{\rm 1D}}^2 |g_{\mathrm{1D}}|^2 }{ \hbar^2} \, . \,
\label{eqn:mis1} \ee The maximum growth rate \be
\gamma_{\mathrm{mg}}= 2 \omega_{\rho}\, |a|\,  n_{{\rm 1D}}
\label{eqn:growth} \ee is obtained at wavenumber \be
k_{\mathrm{mg}}=1/ \xi \, ,\,\, \xi \equiv \ell_{\rho}/\sqrt{4
|a|\,n_{{\rm 1D}}} \, ,\label{eqn:wavenumber}\ee where $\xi$ is
the effective 1D healing length of the
condensate~\cite{akhmediev1992} and $n_{{\rm 1D}} = N|\psi_0|^2 =
N/L$ is the line density. Growth occurs only if $\gamma^2>0$,
which implies $0<k<k_{\mathrm{max}}=\sqrt{2} k_{\mathrm{mg}}$.
This means that nonlinear focusing can only be seeded by
modulations of sufficiently long wavelength and is fastest at the
length scale of $2\pi \xi$.

\begin{figure}[t]
\epsfxsize=8cm \leavevmode  \epsfbox{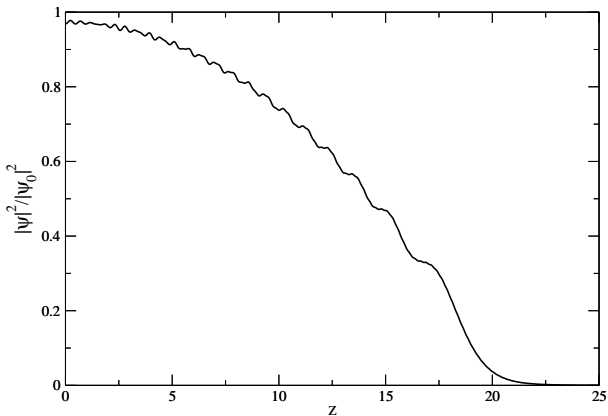}
\caption{\label{fig:3}  Shown are the self-interference fringes of
the order parameter which seed modulational instability, according
to the Feynman propagator for a harmonic oscillator. The initial
density profile was Thomas-Fermi; shown is the ratio of the
density 66 ms later to the original density.  Since the wavelength
of the instability must be on the order of $2\pi \xi$, where $\xi$
is the healing length, solitons form first on the edges of the
cloud, due to the early long wavelength fringes in this region. At
later times the wavelength of the fringes in the center also
becomes longer. Shown is the linear equivalent of the panel
depicting $t=3$ in the full simulation of the 3D Gross-Pitaevskii
equation illustrated in Figs.~\ref{fig:1} and~\ref{fig:2}; all
parameters are the same as the simulation, with length units
scaled to 10 $\mu$m.}
\end{figure}

{ For a non-uniform initial density profile, there are
two ways in which modulational instability can occur.  Either it
can be seeded by noise, or it can be seeded by fringes caused by
self-interference of the order parameter. The time scale of the
two mechanisms turns out to be approximately the same, as shall be
shown in the following. A similar number of solitons results, but
in the former case they form first in the higher density regions
(typically the center, for a Thomas-Fermi-like initial
profile~\cite{dalfovo1}), while in the latter case they form first
at the edges, as illustrated in Figs.~\ref{fig:1} and~\ref{fig:2}
and explained in our previous work~\cite{carr34}.

{ Consider first the case of self-interference.  An
analysis based on the Feynman propagator for the \emph{linear}
Schr\"odinger equation in a harmonic potential shows that
self-interference, or diffraction, of the order parameter leads to
fringes which have the correct length scale to seed modulational
instability~\cite{carr34}.} Our findings of Ref.~\cite{carr34}
were supported later by Kamchatnov {\it et al.}, who used Whitham
theory to describe the nonlinear evolution of the diffraction
pattern of a rectangular initial density
profile~\cite{kamchatnov2003}. Our previous analysis was performed
for a rectangular initial density profile in order to obtain
closed form analytic results~\cite{carr34}. The Feynman propagator
is defined by \be \psi(z,t)=\int dz' \,G(z,t;z',0) \psi(z,0)
\label{eqn:feyn}\, . \ee For a harmonic oscillator, the propagator
is \be G = \frac{\exp \left\{{i}( z^2 - {2zz'}/{\cos \tau} +z'^2)
/ ({2 \ell_{z}^2 \tan \tau}) \right\}} {\ell_{z}\sqrt{2\pi i|\sin
\tau|}} \, , \, \label{eqn:fp2} \ee
where \be\tau \equiv \omega_z t\, .\ee  In the limit $\tau \ll 1$,
and for a rectangular initial density profile, the result of the
integration of Eq.~(\ref{eqn:feyn}) can be Taylor expanded as
\bea &&\left|\psi(z,t)\right|^2/\left|\psi(z,0)\right|^2 \simeq 1+\nonumber\\
&&\sqrt{\frac{8l_z^2\tau}{\pi}}\left[\frac{\sin(k_{+}z+\delta-\frac{\pi}{4})}{L+2z}
\frac{\sin(k_{-}z+\delta-\frac{\pi}{4})}{L-2z}\right]+\nonumber\\
&&\frac{4 l_z^2\tau}{\pi}
\left\{\frac{L^2+4z^2}{(L+2z)^2(L-2z)^2}+\frac{\cos[(k_{+}-k_{-})
z]}{(L+2z)(L-2z)}\right\},\,\,\,\,\:\:\:\:\:\:\label{eqn:fp4} \eea

\be k_{\pm}\equiv \frac{\sec(\tau)z \pm L}{2l_{z}^2\sin(\tau)}\,
,\, \delta\equiv\frac{L^2 \cot(\tau)}{8 l_{z}^2}\,
,\,|z|<\frac{L}{2}\, .\,\label{eqn:fp6} \ee}

{To linear order in $\tau$, the trapping frequency drops out of
the equations, since $l_z^2\tau = \hbar t/m$.
Equations~(\ref{eqn:fp4}) and~(\ref{eqn:fp6}) describe the
formation of fringes.  Note that, according to the argument of the
exponential in the Feynman propagator~(\ref{eqn:fp2}), at the
quarter period the wavefunction is fourier transformed with
respect to its initial state.  Therefore any initial wavefunction
excepting a Gaussian must develop fringes. A time scale can be
estimated from these prefactors in the expansion of
Eq.~(\ref{eqn:fp4}). Fringes appear at a length scale
$\ell_{\mathrm{fringe}}$ at time \be
t\simeq\frac{m}{\hbar}\ell_{\mathrm{fringe}}^2\, .\ee  This
argument can also be made simply by the units in the problem.  The
length scale at which modulational instability is maximally
probable is $\ell_{\mathrm{fringe}}=2\pi \xi$. Therefore, the time
scale for fringe formation leading to modulational instability may
be estimated as \be t_{\mathrm{fringe}} \simeq
\frac{\pi^2}{2\omega_{\rho}|a|\bar{n}_{\mathrm{1D}}}\, ,\ee  where
$\bar{n}_{\mathrm{1D}}$ is the mean linear density and we have
taken the mean density as $\bar{n}=\bar{n}_{\mathrm{1D}}/\pi
\ell_{\rho}^2$ in order to calculate the healing length. For the
parameters of Sec.~\ref{sec:sims},
$t_{\mathrm{interference}}\simeq 41$ ms. This is approximately the
correct time scale, as observed in Figs.~\ref{fig:1}
and~\ref{fig:2}.}

In order to study the problem with a more realistic model than an
initially rectangular density profile, the longitudinal variation
of the density profile is taken as an inverted parabola. This is
characteristic of the Thomas-Fermi limit in a harmonic trap, and
is the generic experimental case~\cite{dalfovo1}.  At the same
time, the transverse wavefunction is taken as a Gaussian, in
keeping with the quasi-1D approximation. The density then takes
the form \bea |\Psi(\vec{r},0)|^2&=&|\psi(z,0)|^2
\frac{1}{\sqrt{\pi}\ell_{\rho}}
\exp\left(-\frac{x^2+y^2}{\ell_{\rho}^2}\right)\label{eqn:tf1}\\
|\psi(z,0)|^2 &=&
\frac{\ell_\rho^2 (R^2 - z^2)}{4 \ell_z^4 |a|} \, ,\label{eqn:tfprofile}\\
R &\equiv& \left(\frac{3 N |a|
\ell_z^4}{\ell_\rho^2}\right)^{1/3}\\
\ell_z&\equiv& \sqrt{\hbar/m|\omega_z|}\, ,\eea where
$N|\psi(z)|^2$ is the longitudinal line density, $R$ is the
Thomas-Fermi radius, and $\ell_z$ is the longitudinal oscillator
length.  The linear development of the wavefunction may be found
at any time by numerical integration of Eq.~(\ref{eqn:feyn}). Note
that, in this case, $\omega_z$ is imaginary for the expulsive
harmonic potential.  An example relevant to Sec.~\ref{sec:sims} is
shown in Fig.~\ref{fig:3}.  The longer wavelength fringes are
clearly visible near the edges of the cloud, as discussed in our
previous work~\cite{carr34}.  This leads to soliton formation near
the edges of the cloud at early times and in the center at late
times. The figure uses the same parameters as the simulations of
Sec.~\ref{sec:sims}, and may be compared to the fourth panel from
the bottom, or $t=3$, in Figs.~\ref{fig:1} and~\ref{fig:2}.

Consider now the case of modulational instability seeded by noise,
rather than interference fringes.  There are two kinds of noise.
They originate in different physical mechanisms. The first is
classical white or coloured noise, which may be induced, for
example, by fluctuations in the trapping potential.  The second is
thermal quantum noise, which corresponds to a thermal distribution
of Boguliubov excitations.  One may estimate the relevance of the
latter from first principles.  The Boguliubov quasi-particle
dispersion relation for a Thomas-Fermi profile is~\cite{dalfovo1}
\be E^{\mathrm{bog}}\equiv
\left\{\sqrt{\displaystyle\frac{\hbar^2k^2}{2m}\left[\frac{\hbar^2k^2}{2m}
+2g N |\Psi(\vec{r})|^2\right]} \quad|r|\leq R
\atop{\displaystyle\frac{\hbar^2k^2}{2m}+\frac{m \omega^2r^2}{2}
-\mu_m \quad\,\,\,\,\,\,\,\,\,\, |r|>R\, .}\right.\,
\label{eqn:boguliubov}\ee  Substituting the wavenumber of maximum
growth for modulational instability, Eq.~(\ref{eqn:wavenumber}),
into Eq.~(\ref{eqn:boguliubov}), the resulting energy is \be
E^{\mathrm{bog}}_{\mathrm{mg}}=\sqrt{3} g \bar{n}\, ,\ee where
$\bar{n}$ is the mean density which can be estimated from
Eq.~(\ref{eqn:tf1}). An experimental situation may, e.g.,
correspond to an initial temperature of the condensate of $T\simeq
T_{\mathrm{BEC}}/2$, with $T_{\mathrm{BEC}}\equiv \hbar
\bar{\omega} [N/\zeta(3)]^{1/3}$ and
$\bar{\omega}\equiv(\omega_{\rho}^2\omega_{z0})^{1/3}$. In this
case, one can estimate the probability of a Boguliubov mode of the
appropriate wavelength to seed modulational instability from the
bosonic number distribution function \be n(E)=\frac{1}{\exp(E/k_B
T)+1}\, .\ee Using the numbers from Sec.~\ref{sec:sims}, one
obtains $T\simeq 0.14 \mu$K and
$E^{\mathrm{bog}}_{\mathrm{mg}}/k_B\simeq 0.23 \mu$K, so that
$n(E^{\mathrm{bog}}_{\mathrm{mg}})\simeq 0.17$. Thus noise caused
by Boguliubov fluctuations is present with a non-negligible
occupation number for the parameters we have chosen.

\begin{figure}[t]
\epsfxsize=8cm \leavevmode  \epsfbox{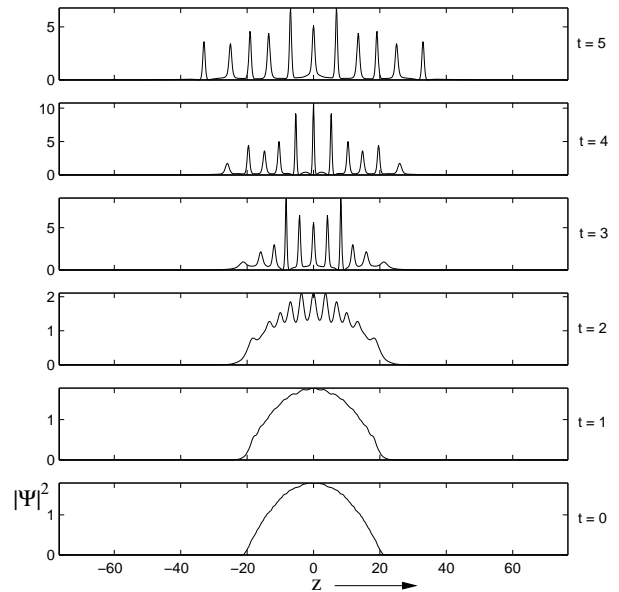}
\caption{\label{fig:4} Shown is the evolution of the density along
a one-dimensional cut at $x,y=0$, with the same parameters as the
simulation of Fig.~\ref{fig:1} but with the addition of noise, as
described in the text.  The time scale is shorter than that
observed in Fig.~\ref{fig:1}, and the solitons form first in the
center of the cloud, rather than on the outside, but the end
result is the same: a set of well-defined solitonic pulses is
evident in the latest (top) panel.  The length and time units are
the same as in Fig.~\ref{fig:1}.}
\end{figure}

A time scale for the growth of seeding fluctuations can be
estimated as \be t_{\mathrm{noise}} \simeq
1/\gamma_{\mathrm{mg}}=\frac{1}{2\omega_{\rho}|a|n_{\mathrm{1D}}}\,
.\ee  This appears to be smaller by a factor of $\pi^2$ than
$t_{\mathrm{fringe}}$.  However, given the qualitative nature of
the two estimates, and the fact that they have the same parameter
dependence, it is not possible to state that noise definitively
dominates over the fringe mechanism.  To test this, we performed
additional simulations with thermally distributed and various
other realizations of noise in the initial condition. These
simulations show that the final result is essentially the same as
that shown in Figs.~\ref{fig:3} and~\ref{fig:1}: 11 solitons
result rather than 13, with formation first in the regions of
higher density and then lower, in contrast to the case of fringes
alone. Figure~(\ref{fig:4}) illustrates an example of the early
time evolution of the density.  Solitons begin to form at $t\simeq
30$ ms, in contrast to Fig.~\ref{fig:1}, where fringes appear to
self-focus at $t\simeq 40$ ms.  We conclude that the two
mechanisms do indeed coexist.

Noise was added into the simulations in the following manner.
Because the longitudinal and transverse degrees of freedom are
represented differently in the Laguerre DVR cylindrically
symmetric algorithm we used, and most of the grid is not occupied
in the initial state, one generates a great deal of high energy
and high frequency oscillations with usual noise schemes, such as
adding a small random number to the wavefunction or its Fourier
transform. The fluctuations that would seed modulational
instability are of long wavelength. Therefore, starting from the
wavefunction on the grid in position space an FFT was implemented
in the longitudinal direction. Then the wavefunction was
multiplied at each point by $1+n\,r$, where $r$ was a random
number between $-0.5$ and $0.5$ and $n$ was the noise level.  In
Fig.~\ref{fig:4}, $r=0.1$ was used.  In the transverse direction,
the noise was added on only half the grid closest to the center.
Finally, in order to allow the noise to ``thermalize'' as much as
possible, the wavefunction was propagated in real time for
positive scattering length, until the noise had fully distributed
itself, {\it i.e.}, for times much greater than $2\pi/|\omega_z|$.
This was intended to represent, qualitatively, a semi-classical
approximation to a thermal
distribution~\cite{sinatra2001,davis2002a} of Boguliubov modes.
Figure~\ref{fig:4} then follows the real time evolution starting
with this initial wavefunction after the scattering length is
turned negative and the trap is changed to be longitudinally
expulsive. In simulations with smaller noise levels we observe a
coexistence regime and crossover of both seeding mechanisms, as
the growth of thermal fluctuations is significantly delayed when
they are initially very small.

\subsection{Phase decoherence time}
\label{ssec:decohere}

A condensate adiabatically split into two halves on a time scale much shorter
than the quantum revival time has an initially well-defined relative
phase~\cite{wright1996,menotti2001}.  Estimates for the decoherence time~\cite{lewenstein1996}
have been made in a number of specific contexts in the literature, as
for example in the two-well problem~\cite{javanainen1997,castin1997}
or for two spin states~\cite{sinatra2000}.  A general discussion of this
issue may be found in Ref.~\cite{cct1999}.  Here, we follow the straightforward estimates
made in a recent article on atom interferometers using Bose-Einstein condensates,
in which the phase decoherence time was studied experimentally~\cite{shin2004,leanhardtcommunication2004}.

The Schr\"odinger phase of each soliton may be estimated from its
wavefunction, which is proportional to $\exp(-i \mu t/\hbar)$: \be
\Delta \phi = t\Delta \mu /\hbar \, , \label{eqn:phase1}\ee  where
$t$ is the decoherence time. The chemical potential may be
determined from Eq.~(\ref{eqn:gpe1d}), {\it i.e.}, in the quasi-1D
approximation, to be \be \mu = \frac{1}{2}\hbar \omega_{\rho}
\left(\frac{N}{N_s}\right)^2
\left(\frac{a}{\ell_{\rho}}\right)^2\, ,\label{eqn:phase2}\ee
where it has been assumed that $N_s$ solitons of equal amplitude
are formed.  Then, from the derivative of Eq.~(\ref{eqn:phase2})
with respect to $N$, \be \frac{\Delta\mu}{\Delta N}\simeq
\frac{2\mu}{N}\, .\label{eqn:phase2.5}\ee  For Poissonian number
fluctuations, one may take $\Delta N = \sqrt{N}$. Setting
$\Delta\phi = 2\pi$, which is a measure of complete uncertainty in
the relative phase and therefore decoherence,
Eqs.~(\ref{eqn:phase1}) and~(\ref{eqn:phase2.5}) yield \be t =
\frac{\pi \hbar \sqrt{N}}{|\mu|} \, . \label{eqn:phase3}\ee
Substituting Eq.~(\ref{eqn:phase2}) into Eq.~(\ref{eqn:phase3}),
\be t \simeq 2 \left(\frac{N_s}{N}\right)^2 \sqrt{N}
\left(\frac{\ell_{\rho}}{a}\right)^2 \frac{1}{\omega_{\rho}}\,
.\label{eqn:phase4}\ee We note that, in contrast to a repulsive
condensate in Thomas-Fermi limit, for which the decoherence time
$t \propto N^{1/10}$, in the case of solitonic pulses formed by
modulational instability $t \propto N^{-3/2}$.  However, unlike in
the repulsive case, the number of atoms is limited by the collapse
conditions of Secs.~\ref{ssec:2dcollapse}
and~\ref{ssec:collapse3d}.  { For the parameters of
Sec.~\ref{sec:sims}, the phase decoherence time may be calculated
to be about 540 ms, so that the solitonic pulses shown in the
figures are expected to be coherent over the evolution period
depicted.}

\section{Discussion of simulations: number of solitons and refined stability conditions}
\label{sec:params}

In experiments, a { good} model of the initial state
of the condensate when the scattering length is changed from
positive to negative is a longitudinal Thomas-Fermi density
profile~\cite{dalfovo1}. In the following, explicit estimates for
the number of solitons generated by such a profile and criteria to
avoid collapse, in terms of the parameters of a possible
experiment, is compared to the more idealized situation discussed
in { Sec.~\ref{sec:criteria}.}

Under the condition that a suitable seed for the modulational
instability is provided, one can estimate the number of solitons
generated for an initially homogeneous profile along the $z$
direction of length $L$ by \be \label{eqn:Nshom}
   N_s^{{\rm hom}} \sim \frac{L}{2 \pi \xi}  = \sqrt{\frac{N |a|\; L}{\pi
   \ell_\rho^2}}\, ,
\ee where the modulational instability is assumed to take place at
the wavelength of maximum growth given by
Eq.~(\ref{eqn:wavenumber}).

The 2D collapse criterion for the initial state (\ref{eqn:transv})
can be refined by demanding that the solitons formed by
modulational instability are themselves stable against 3D collapse
and satisfy Eq.~(\ref{eqn:collapse3d}). For simplicity, it is
assumed that the initial condensate is split up into $N_s$
solitons of equal amplitude.  As seen from the numerical
simulations { of Secs.~\ref{sec:sims} and~\ref{ssec:interfere},}
this is not strictly true, but it serves as a useful order of
magnitude estimate. For the homogeneous initial profile, one finds
from Eq.~(\ref{eqn:Nshom}) \be 8 \pi  |a| N/L < \frac{8}{\pi}
(\eta_c^{{\rm 3 D}})^2 = 1.0\ldots \, . \ee This estimate assumes
a quasi-1D initial state, where the transverse trapping is tight,
so that $|a|\ll \ell_\rho \ll \xi$. Note that under these
conditions the transverse oscillator length $\ell_\rho$ does not
enter the collapse criteria for homogeneous initial density
profiles.

In the case of an inhomogeneous initial density profile the above
estimates can be generalized by assuming that the length scale of
$2\pi\xi$ for the modulational instability is still valid locally.
The number of solitons can thus be estimated as \be N_s = \int
\frac{dz}{2 \pi \xi(z)}\, . \ee For the Thomas-Fermi density
profile of Eq.~\ref{eqn:tfprofile},
 one obtains \be \label{eqn:NsTF}
  N_s^{{\rm TF}} = \frac{1}{2} \left(\frac{3 N |a|\;
     \ell_z}{\ell_\rho^2}\right)^{\frac{2}{3}} \,.
\ee

Similarly, one can derive a collapse criterion for the
Thomas-Fermi case.  The aspect ratio of the trap enters
explicitly, and the criterion to avoid collapse is given by \be
\label{ieq:TFcollapse} \frac{|a| N}{\ell_z} < \frac{(\eta_c^{{\rm
3 D}})^3}{576} \frac{\ell_z}{\ell_\rho} \approx 0.0347
\frac{\ell_z}{\ell_\rho}. \ee These estimates restrict the number
of particles for a given scattering length that can be used in a
particular trap geometry. Combining the above results, one finds
an upper bound for the number of solitons that can be generated
from a given trap geometry used for preparing the initial state.
In the case of a rectangular initial state, or homogeneous case,
one obtains \be \label{eqn:Nshomest}
  N_s^{{\rm hom}} < 0.0635 \frac{L}{\ell_\rho}\, .
\ee For a condensate initially described by a Thomas-Fermi profile
one finds \be \label{eqn:NsTFest}
  N_s^{{\rm TF}} < 0.055 \left(\frac{\ell_z}{\ell_\rho}\right)^2 \, .
\ee The bounds (\ref{eqn:Nshomest}) and (\ref{eqn:NsTFest}) were
based on the 3D collapse criterion of Eq.~(\ref{eqn:collapse3d}).
A similar analysis based on the 2D criterion (\ref{eqn:transv})
for the initial wavefunction yields the same scaling but a
prefactor which is an order of magnitude larger, {\it i.e.}, a
less stringent constraint.

One may now compare these refined estimates to the simulations of
Secs.~\ref{sec:sims} and~\ref{ssec:interaction}. First, the choice
of $-a N / \ell_{z0}$ was taken to be about one order of magitude
smaller than the upper limit given by the criteria to avoid
collapse, according to Eq.~(\ref{ieq:TFcollapse}). Secondly,
Eq.~(\ref{eqn:NsTF}) predicts an upper bound on the number of
solitons to be $N_s \approx 13.3$; in the simulations, between 11
and 14 solitons were observed, depending on the noise level.
Thirdly, all forms of collapse have been successfully avoided,
including soliton--soliton interactions which might lead to
secondary collapse.

Many other parameter regimes were studied numerically.  It was
found that, for a rectangular initial profile { and no noise}, as
was studied analytically in Ref.~\cite{carr34}, increasing the
strength of the nonlinearity to the critical value of $\eta_{{\rm
hom}}=1$ (see Eq.~(\ref{eqn:Nshomest}) below) brought about
immediate collapse at the \emph{borders} of the condensate.  That
is, the first soliton formed collapsed.  An order of magnitude
decrease in $\eta_{{\rm hom}}$ to 0.1 led to delayed collapse
which occurred after all solitons had been formed, while for
$\eta_{{\rm hom}}=0.01$ no collapse occurred. Note that a
rectangular initial density profile may be created by optically
induced potentials which form end-caps~\cite{bongs1}, as were used
in the experiment of Ref.~\cite{strecker1}.

\section{Conclusion}
\label{sec:conclude}

We have shown both numerically and analytically that a pulsed
atomic soliton laser is viable.  In particular, the figures
illustrate the evolution of such an atom laser with a set of
realistic parameters that could be realized in straightforward
adaptions of existing BEC apparatuses~\cite{carr29,strecker1}. It
was shown that all phenomena leading to instability, namely,
two-dimensional primary collapse, three-dimensional primary
collapse, explosion of individual solitonic pulses brought about
by the longitudinally expulsive harmonic trapping potential, and
secondary collapse caused by soliton--soliton interaction, could
be avoided by the proper choice of parameters.  Typical parameters
were $10^4$ particles, a final scattering length of $a\sim -3
a_0$, and trapping frequencies on the order of $2\pi \times 2.2$
kHz by $2\pi\times 2.2$ kHz by $2\pi i \times 2.5$ Hz. After
formation via modulational instability seeded by { a
combination of self-interference of the condensate order parameter
and noise due to the presence of Boguliubov quasiparticles and
fluctuations in the trapping potential,} propagating solitonic
pulses self-cool to $T=0$ on a time scale of $1/|\omega_z|$
through the emission of a fraction of a percent of the total
number of particles~\cite{satsuma1,carr30}.

{ In most previous work on attractive Bose-Einstein
condensates, regimes or cycles of runaway instability were
explored~\cite{sackett2,donley1}.  Even in the cases where a
stable final state was produced, as for example in
Refs.~\cite{carr29,strecker1}, the majority of the atoms were lost
to collapse.  In contrast, we have here suggested a way to avoid
collapse entirely and take advantage of the instabilities inherent
in switching the interactions in a BEC from repulsive to
attractive to produce a useful device: namely, a pulsed atomic
soliton laser.}

We acknowledge useful discussions with Jinx Cooper, {
Simon Gardiner}, and Murray Holland, and thank H.-D.~Meyer for a
preprint of Ref.~\cite{ref:laguDVR} prior to publication.
L.~D.~Carr gratefully acknowledges the support of the National
Science Foundation via grant no.~MPS-DRF 0104447 and the U.S.
Department of Energy, Office of Basic Energy Sciences via the
Chemical Sciences, Geosciences and Biosciences Division.


\begin{thebibliography}{10}

\bibitem{hasegawa1}
A. Hasegawa, {\em Optical Solitons in Fibers} (Springer-Verlag,
New York,
  1990).

\bibitem{drazin1}
P.~G. Drazin and R.~S. Johnson, {\em Solitons: an Introduction}
(Cambridge
  Univ. Press, Cambridge, 1989).

\bibitem{agrawal1}
G.~P. Agrawal, {\em Nonlinear Fiber Optics}, 2nd ed. (Academic
Press, San
  Diego, 1995).

\bibitem{kivshar3}
Y.~S. Kivshar, Physics Reports {\bf 298},  81  (1998).

\bibitem{sulem1}
C. Sulem and P.~L. Sulem, {\em Nonlinear Schr\"odinger Equations:
Self-focusing
  Instability and Wave Collapse} (Springer-Verlag, New York, 1999).

\bibitem{mewes1}
M.~O. Mewes, M.~R. Andrews, D.~M. Kurn, D.~S. Durfee, C.~G.
Townsend, and W.
  Ketterle, Phys. Rev. Lett. {\bf 78},  582  (1997).

\bibitem{hagley1}
E.~W. Hagley, L. Deng, M. Kozuma, J. Wen, K. Helemerson, S.~L.
Rolston, and
  W.~D. Phillips, Science {\bf 283},  1706  (1999).

\bibitem{anderson1998}
B.~P. Anderson and M.~A. Kasevich, Science {\bf 282},  1686
(1998).

\bibitem{bloch2}
I. Bloch, T.~W. H\"ansch, and T. Esslinger, Phys. Rev. Lett. {\bf
82},  3008
  (2000).

\bibitem{leanhardt2002}
A.~E. Leanhardt, Y. Shin, A.~P. Chikkatur, D. Kielpinski, W.
Ketterle, and
  D.~E. Pritchard, Phys. Rev. Lett. {\bf 90},  100404  (2003).

\bibitem{ott2001}
H. Ott, J. Fortagh, G. Schlotterbeck, A. Grossmann, and C.
Zimmermann, Phys.
  Rev. Lett. {\bf 87},  230401  (2001).

\bibitem{hansel2001}
W. H\"ansel, P. Hommelhoff, T.~W. H\"ansch, and J. Reichel, Nature
{\bf 413},
  498  (2001).

\bibitem{kasevich2002}
M.~A. Kasevich, Science {\bf 298},  1363  (2002).

\bibitem{sauer2001}
J.~A. Sauer, M.~D. Barrett, and M.~S. Chapman, Phys. Rev. Lett.
{\bf 87},
  270401  (2001).

\bibitem{satsuma1}
J. Satsuma and N. Yajima, Prog. of Theor. Phys. (Suppl.) {\bf 55},
284
  (1974).

\bibitem{carr30}
L.~D. Carr and Y. Castin, Phys. Rev. A {\bf 66},  063602  (2002).

\bibitem{dalfovo1}
F. Dalfovo, S. Giorgini, L.~P. Pitaevskii, and S. Stringari, Rev.
Mod. Phys.
  {\bf 71},  463  (1999).

\bibitem{carr29}
L. Khaykovich, F. Schreck, F. Ferrari, T. Bourdel, J. Cubizolles,
L.~D. Carr,
  Y. Castin, and C. Salomon, Science {\bf 296},  1290  (2002).

\bibitem{strecker1}
K.~E. Strecker, G.~B. Partridge, A.~G. Truscott, and R.~G. Hulet,
Nature {\bf
  417},  150  (2002).

\bibitem{ruprecht1}
P.~A. Ruprecht, M.~J. Holland, K. Burnett, and M. Edwards, Phys.
Rev. A {\bf
  51},  4704  (1995).

\bibitem{vogels1}
J.~M. Vogels, C.~C. Tsai, R.~S. Freeland, S.~J. J. M.~F.
Kokkelmans, B.~J.
  Verhaar, and D.~J. Heinzen, Phys. Rev. A {\bf 56},  R1067  (1997).

\bibitem{inouye1}
S. Inouye, M.~R. Andrews, J. Stenger, H.-J. Miesner, D.~M.
Stamper-Kurn, and W.
  Ketterle, Nature {\bf 392},  151  (1998).

\bibitem{abdullaev2003}
F.~K. Abdullaev, J.~G. Caputo, R.~A. Kraenkel, and B.~A. Malomed,
Phys. Rev. A
  {\bf 67},  013605  (2003).

\bibitem{carr34}
L.~D. Carr and J. Brand, Phys. Rev. Lett. {\bf 92},  040401
(2004).

\bibitem{hasegawa2}
A. Hasegawa and W.~F. Brinkman, IEEE J. Quantum Electron. {\bf
16},  694
  (1980).

\bibitem{sinatra2001}
A. Sinatra, C. Lobo, and Y. Castin, Phys. Rev. Lett. {\bf 87},
210404 (2001).

\bibitem{davis2002a}
M.~J. Davis, S.~A. Morgan, and K. Burnett, Phys. Rev. A {\bf 66},
053618
  (2002).

\bibitem{ref:laguDVR}
{For representing the transverse degree of freedom a Laguerre DVR
was used as
  described in C.~W.~McCurdy, W.~A.~Isaacs, H.-D.~Meyer, and T.~N.~Rescigno},
  Phys. Rev. A {\bf 67},  042708  (2003).

\bibitem{weinstein1983}
M.~I. Weinstein, Comm. Math. Phys. {\bf 87},  567  (1983).

\bibitem{pitaevskii1996}
L.~P. Pitaevskii, Phys. Lett. A {\bf 221},  14  (1996).

\bibitem{olshanii1}
M. Olshanii, Phys. Rev. Lett. {\bf 81},  938  (1998).

\bibitem{gammal1}
A. Gammal, T. Frederico, and L. Tomio, Phys. Rev. A {\bf 64},
055602  (2001).

\bibitem{ng2004}
H.~K. Ng, K.~D. Moll, and A.~L. Gaeta, 2004, submitted to Phys.
Rev. A.

\bibitem{cct1999}
C. Cohen-Tannoudji, Cours de physique atomique et mol\'eculaire,
  http://www.lkb.ens.fr/$\sim$cct/cours/, 1999.

\bibitem{michinel1}
H. Michinel, V. P\'erez-Garc\'ia, and R. de~la Fuente, Phys. Rev.
A {\bf
  60},  1513  (1999).

\bibitem{gordon2}
J.~P. Gordon and H.~A. Haus, Opt. Lett. {\bf 11},  665  (1986).

\bibitem{elyutin1}
P.~V. Elyutin, A.~V. Buryak, V.~V. Gubernov, R.~A. Sammut, and
I.~N. Towers,
  Phys. Rev. E {\bf 64},  016607  (2002).

\bibitem{khawaja2002}
U.~A. Khawaja, H.~T.~C. Stoof, R.~G. Hulet, K.~E. Strecker, and
G.~B.
  Partridge, Phys. Rev. Lett. {\bf 89},  200404  (2002).

\bibitem{carr22}
L.~D. Carr, M.~A. Leung, and W.~P. Reinhardt, J. Phys. B: At. Mol.
Opt. Phys.
  {\bf 33},  3983  (2000).

\bibitem{fortagh2002}
J. Fortagh, H. Ott, S. Kraft, A. Gunther, and C. Zimmermann, Phys.
Rev. A {\bf
  66},  041604  (2002).

\bibitem{garciaripoll2}
J.~J. Garc\'ia-Ripoll, V.~M. P\'erez-Garc\'ia, and V. Vekslerchik,
Phys. Rev. E
  {\bf 64},  056602  (2001).

\bibitem{akhmediev1992}
N.~N. Akhmediev, V.~I. Korneev, and R.~F. Nabiev, Opt. Lett. {\bf
17},  393
  (1992).

\bibitem{kamchatnov2003}
A.~M. Kamchatnov, A. Gammal, F.~K. Abdullaev, and R.~A. Kraenkel,
Phys. Lett. A
  {\bf 319},  406  (2003).

\bibitem{wright1996}
E.~M. Wright, D.~F. Walls, and J.~C. Garrison, Phys. Rev. Lett.
{\bf 77},  2158
   (1996).

\bibitem{menotti2001}
C. Menotti, J.~R. Anglin, J.~I. Cirac, and P. Zoller, Phys. Rev. A
{\bf 63},
  023601  (2001).

\bibitem{lewenstein1996}
M. Lewenstein and L. You, Phys. Rev. Lett. {\bf 77},  3489
(1996).

\bibitem{javanainen1997}
J. Javanainen and M. Wilkens, Phys. Rev. Lett. {\bf 78},  4675
(1997).

\bibitem{castin1997}
Y. Castin and J. Dalibard, Phys. Rev. A {\bf 55},  4330  (1997).

\bibitem{sinatra2000}
A. Sinatra and Y. Castin, Eur. Phys. J. D {\bf 9},  319  (2004).

\bibitem{shin2004}
Y. Shin, M. Saba, T.~A. Pasquini, W. Ketterle, D.~E. Pritchard,
and A.~E.
  Leanhardt, Phys. Rev. Lett. {\bf 92},  050405  (2004).

\bibitem{leanhardtcommunication2004}
2004, private communication, A. E. Leanhardt, MIT.

\bibitem{bongs1}
K. Bongs, S. Burger, S. Dettmer, D. Hellweg, J. Arlt, W. Ertmer,
and K.
  Sengstock, Phys. Rev. A {\bf 63},  031602  (2001), e-print cond-mat/0007381.

\bibitem{sackett2}
C.~A. Sackett, H.~T.~C. Stoof, and R.~G. Hulet, Phys. Rev. Lett.
{\bf 80},
  2031  (1998).

\bibitem{donley1}
E.~A. Donley, N.~R. Claussen, S.~L. Cornish, J.~L. Roberts, E.~A.
Cornell, and
  C.~E. Wieman, Nature {\bf 412},  295  (2001).

\end{thebibliography}

\end{document}